\documentclass[11pt]{article}
\pdfoutput=1

\usepackage[pdftex]{graphicx}

\usepackage{import} 
\usepackage{color} 
\usepackage{transparent}

\usepackage{afterpage}

\graphicspath{{cavifig/}}

\title{Modeling high impedance connecting links and cables below \mbox{1 Hz}}

\author{Andrea Giachero$^{a,b}$, Claudio Gotti$^{a,c}$\thanks{Corresponding author},\\
Matteo Maino$^{a,b}$, Gianluigi Pessina$^{a,b}$\\
\\
\llap{$^a$}Sezione di Milano Bicocca, INFN,\\Piazza della Scienza 3, 20126, Milano, Italy\\
\\
\llap{$^b$}Dipartimento di Fisica G. Occhialini,\\Universit\`a degli Studi di Milano Bicocca,\\
\\
Piazza della Scienza 3, 20126, Milano, Italy\\
\llap{$^c$}Dipartimento di Elettronica e Telecomunicazioni,\\Universit\`a degli Studi di Firenze,\\
via S. Marta 3, 50139, Firenze, Italy\\
\\
 E-mail: claudio.gotti@mib.infn.it}

\begin{document}

\maketitle


\begin{abstract}
High impedance connecting links and cables are modeled at low frequency in terms of their impedance to ground and to neigbouring connecting links. The impedance is usually considered to be the parallel combination of a resistance and a capacitance. While this model is adequate at moderate and low frequency, it proved to be not satisfactory at very low frequency, in the fractions of Hz range. Deep characterization was carried out on some samples down to 10 $\mu$Hz, showing that an additional contribution to capacitance can emerge. A model was developed to explain and account for this additional contribution.
\end{abstract}

\section{Introduction}
The quality of the connecting links and cables is most often a crucial aspect of electric and electronic systems. At low frequency (from the audio range down to DC), where the inductive contributions are negligible, the connecting links are characterized in terms of their series resistance and of their parallel conductance and capacitance to ground or to other neighbouring conductors. These are the main electrical quantities which must be considered in the evaluation of the quality of a connecting link and of its compliance with the performance of the system.

In this paper we are concerned with high impedance signal links, carrying low frequency voltage signals from a capacitive or resistive source to an ideal voltage amplifier with infinite input impedance. This excludes effects due to high current flow and power dissipation in the links. Examples are the read out of the CUORE \cite{cuore1, cuore2} and Lucifer \cite{lucifer1} experiments, where thermistors inside a cryostat, whose impedance is in the G$\Omega$ range, are connected to the front end amplifiers at room temperature through high impedance connecting links a few meters long. The signal bandwidth in this case extends from DC up to about 100 Hz.

In describing the modelization of connecting links, we assume each cable to be individually shielded, so that the dominant parasitic contributions are between each link and ground. The parasitic components to neighbouring connecting links will not be considered in the following, but all the models could be straightforwardly extended to account for them.

\section{The basic model}
\begin{figure}[ht]
\centering
 \def\svgwidth{150px}
 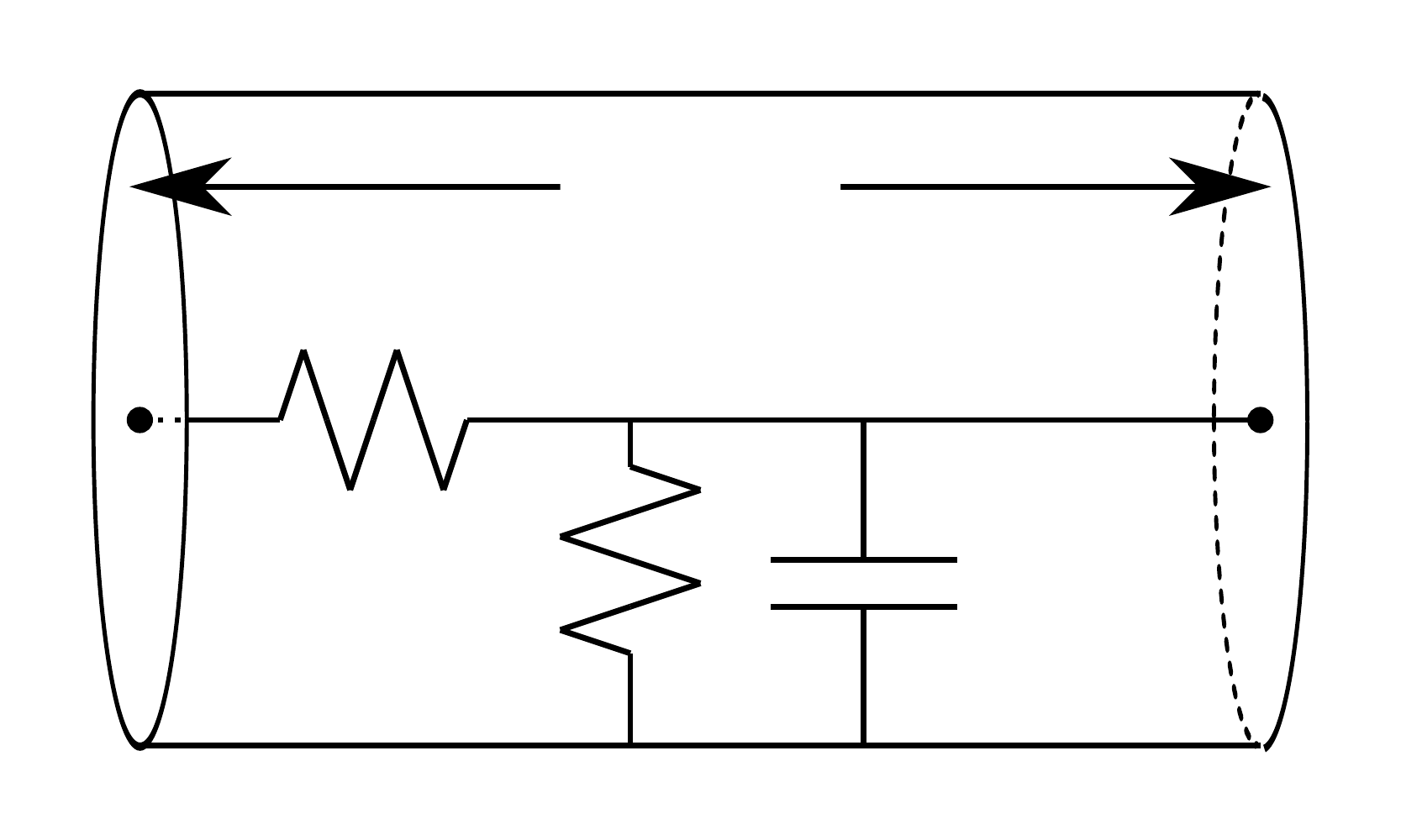
 \hspace{0.1\linewidth}
 \def\svgwidth{150px}
 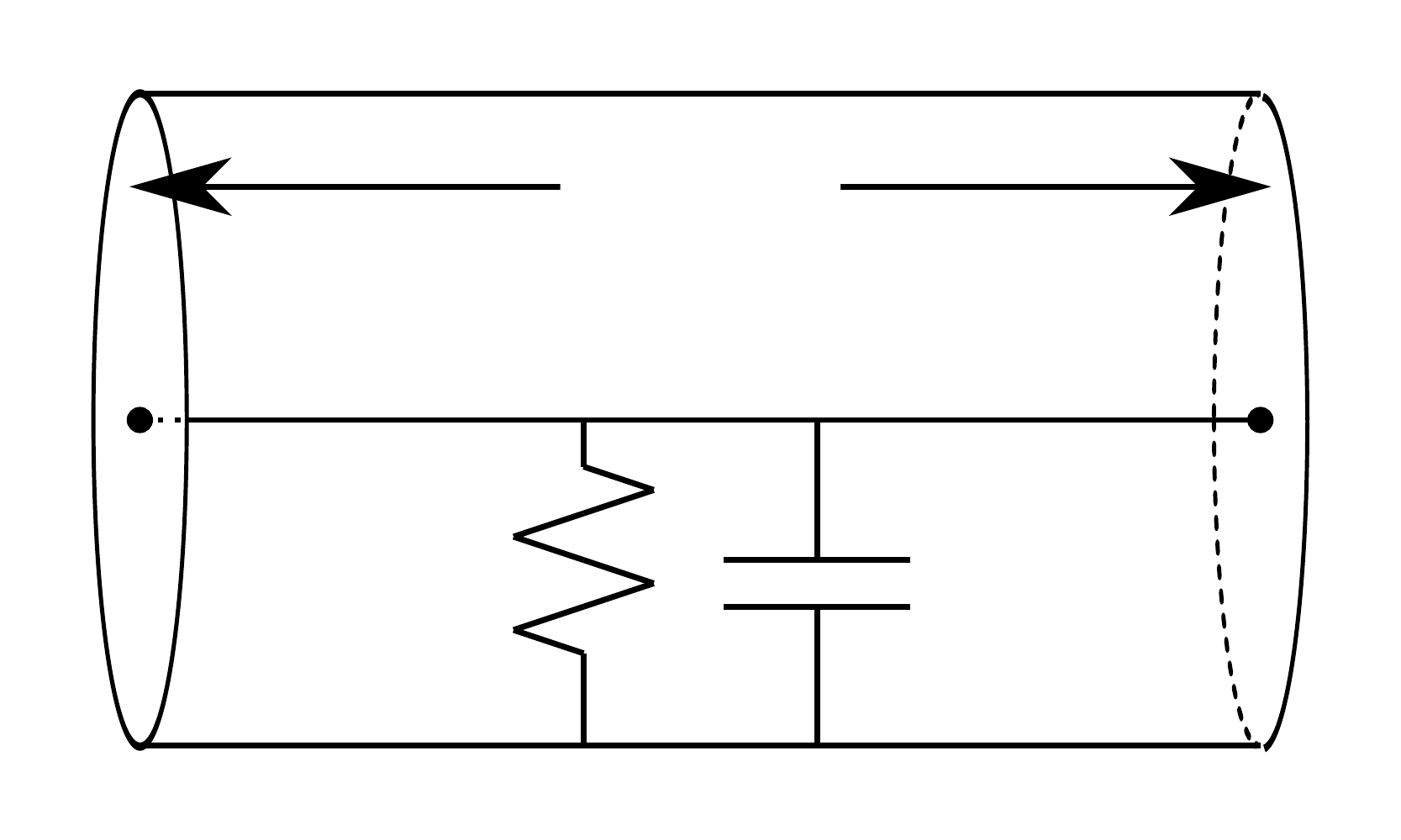
\caption{On the left hand side, a model commonly used to describe the moderate and low frequency behaviour of a cable element of length $dx$ is shown. On the right hand side, the model was simplified neglecting the series resistance.}
\label{basicmodel}
\end{figure}

According to the basic model, each section of the cable of length $dx$ can be modeled with the three impedances $R(x)$, $G(x)$ and $C(x)$, accounting for the series resistance, parallel parasitic conductance and parallel capacitance respectively. The quantities $R(x)$, $G(x)$, $C(x)$ are considered per unit length, and if the cable is not homogeneous their values depend on their position $x$ along the cable, going from $x=0$ to $x=L$, where $L$ is the length of the cable. This model is depicted in the left hand side of figure \ref{basicmodel} for a cable element of length $dx$; integrating over the cable length from $x=0$ to $x=L$ gives the impedance of the entire cable.

For the moment, let us consider the cable to be homogeneous, and drop the dependence of $R$, $G$ and $C$ from the position $x$ ($R$, $G$ and $C$ are still considered per unit length). There are two time constants involved: $R C L^2$ gives the order of magnitude of the low pass effect for a cable of length $L$, while $C / G$ gives the order of magnitude of the boundary between the DC regime, where $G$ dominates over $C$, and the AC regime, where $C$ dominates over $G$. This second time constant does not depend on $L$. Unless the cable is extremely long, extremely thin, or of a very poor quality, the first time constant is smaller than the second of several order of magnitudes. We are here interested in the very low frequency behaviour of the connecting link, and we will thus neglect $R$ in the following. The paper will focus on the model depicted in the right hand side of figure \ref{basicmodel}, probing its adequacy in describing the connecting link on the time scale of  $C / G$, and extending it to account for measured deviations.


\section{Measurement setup}

\begin{figure}[ht]
\centering
 \def\svgwidth{300px}
 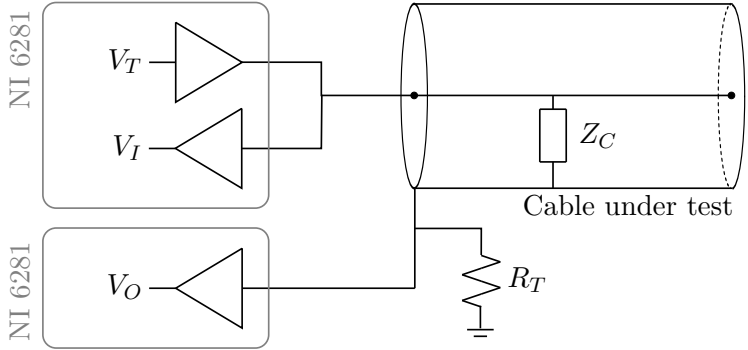
\caption{The setup for cable characterization.}
\label{setup}
\end{figure}

The setup used for a frequency dependent measurement of the cable is shown in figure \ref{setup}. A National Instruments 6281 card was used to generate a test sinusoidal signal $V_I$, with settable frequency. The test signal was applied at one end of the cable, while the other end was left open. The shielding of the cable was connected to ground through a resistor $R_T$. The cable was held at the constant temperature of 30 $^\circ$C inside a V\"{o}tsch VT7004 climatic chamber. The same NI 6281 which generates the test signal was used to acquire it, while another NI 6281 was used to acquire the signal $V_O$ on the shielding. The reason why two NI 6281 cards were used instead of one is to avoid switching the same input between $V_O$ and $V_I$, which would affect the measurement of $V_O$, which is at high impedance.

For each frequency value between 10 $\mu$Hz and 10 Hz, five periods of the test sine wave were applied to the cable and acquired. The first two are necessary to allow the transients to disappear. The last three periods were fitted with sine waves, extracting the amplitude and phase of $V_O$ and $V_I$.
The transfer function of the setup is
\begin{equation}
\frac{V_O}{V_I} = \frac{R_T}{Z_C (s) + R_T},
\end{equation}
where $Z_C$ is the impedance of the cable.
It is clear that if $R_T$ is much larger or much smaller than $Z_C$, the setup cannot achieve a good sensitivity. For this reason, two values for $R_T$ were used in the characterization: 10 G$\Omega$ for the lower part of the spectrum, where $Z_C$ is larger, and 1 M$\Omega$ for the upper part of the spectrum, where $Z_C$ is smaller.
The whole spectrum of $Z_C$ versus frequency was then reconstructed over a wide frequency range.

The fit of the measured $Z_C$ was performed by considering both its amplitude and phase at the same time, by minimizing the distance in the complex plane given by
\begin{equation}
D_ p= \sum_i \frac{\left|y_i - f_p(x_i) \right|}{\left|y_i\right|},
\label{complexdist}
\end{equation}
as a function of the fit parameters $p$. Here $x_i$ are the frequency values, $y_i$ the measured (complex) values of the transfer function at each frequency $x_i$, and $f_p$ is the fitting function, which depends on the fit parameters $p$. The complex distance defined in \ref{complexdist} is normalized by the module of $y_i$: in this way the fit is not biased towards the points where the sampled function is larger.

\section{Limits of the basic model}

For a given cable of length $L$, the basic model at the right of figure \ref{basicmodel} gives the following expression for the cable impedance $Z_C$:
\begin{equation}
Z_C^{-1} =  \int_0^L \left[ G(x) + sC(x) \right] dx =G_C + s C_C,
\label{basicimp}
\end{equation}
where $G_C = \int_0^L G (x) dx$ is the total conductance of the cable, and $C_C =  \int_0^L C (x) dx$ is its total capacitance.

\begin{figure}[ht]
\centering
 \includegraphics[scale=0.8]{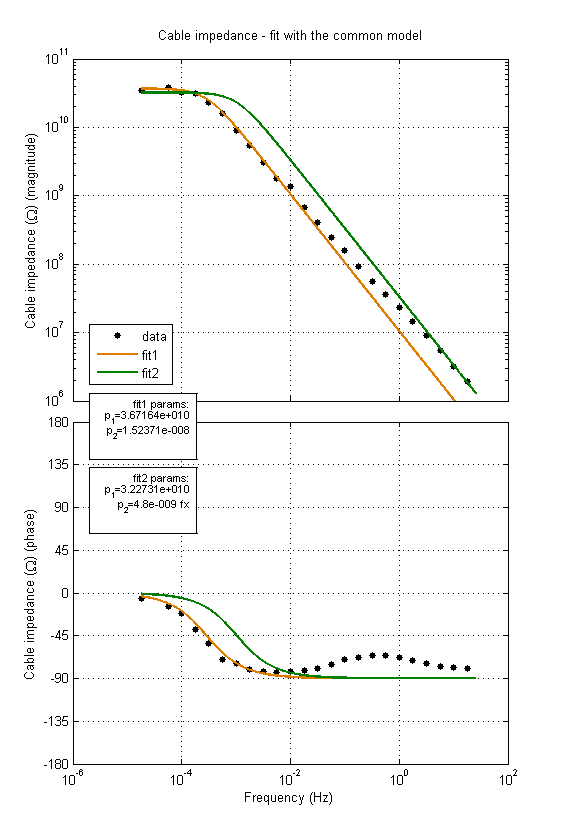}
\caption{The measured cable impedance fitted with the basic model with different capacitance values.}
\label{basicfit}
\end{figure}

Figure \ref{basicfit} shows the plot of the impedance versus frequency for a cable about four meters long, made of 13 differential pairs, connected together giving an equivalent cable made of only 2 conductors. One of the two conductors is considered to be the shielding, just for clarity. In this way, the parasitic conductance and capacitance of each pair of the cable appear multiplied by 13. The low frequency characterization extends down to 10 $\mu$Hz.
The data were fitted with the function given by \ref{basicimp}, that is, rearranging the terms,
\begin{equation}
f_p = \frac{ p_1 }{1+s p_1 p_2 },
\label{fitfunc1}
\end{equation}
where $p_1$ = $G_C^{-1}$ and $p_2$ = $C_C$.
The plot shows in orange the best fitting curve obtained with the basic model: the fit gives $p_1$ = 36.7 G$\Omega$ and $p_2$ = 15.2 nF. 
It is clear looking at figure \ref{basicfit} that with these values there is no agreement between the data and the basic model in the higher portion of the spectrum.
Moreover, the capacitance value obtained with this fit does not match the value of capacitance that is seen at high frequency by measuring the cable with a capacitance meter, which is 4.8 nF.

In the same plot, green curves, the fit was forced with $C_C$ = 4.8 nF, to match the capacitance value measured at high frequency with the impedance meter.
In this case, the fitting curve matches the higher frequency values of the measured curve, but there is a large discrepancy in the middle range. It is then clear that the basic model, with only the two parameters $p_1$ = $G_C^{-1}$ and $p_2$ = $C_C$, fails to accurately describe the behaviour of the cable impedance in the very low frequency range between DC and 1 Hz.

\afterpage{\clearpage}

\section{The extended model}

The discrepancy between the measured data and the basic model, together with the fact that the measurement with the capacitance meter gives a capacitance value which is about 1/3 of the value which comes from the fit with the basic model, suggest the need to extend the basic model.
The main feature of an extended model would need to be the presence of an additional capacitance that is hidden in the higher portion of the spectrum, and becomes relevant only below 1 Hz.

\begin{figure}[ht]
\centering
 \def\svgwidth{300px}
 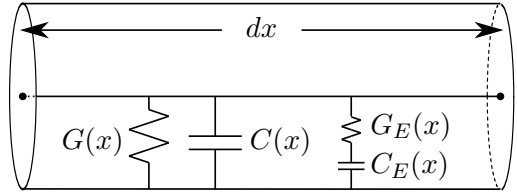
\caption{The extended model for a cable section of elementary length $dx$.}
\label{extmodel}
\end{figure}

The elementary cable element of length $dx$ can be modified to account for the additional capacitance, as depicted in figure \ref{extmodel}.
An additional capacitance was added, whose value is hidden at high frequency because of the resistance it has in series.
The physical motivation for this is not hard to accept: since the capacitance between two electrodes increases with decreasing distance, and since there is a conductive path between the inner and outer conductors of the cable, modeled by $G(x)$, it is reasonable that an intermediate contribution can arise, given by the series combination of a conductive path $G_E(x)$ and a capacitance $C_E(x)$.
The value of $C_E(x)$ is expected to be larger than $C(x)$, since the distance between its electrodes is smaller, and the value of $G_E(x)$ is expected to be larger than $G(x)$, since it represents only a fraction of the conductive path between the inner and outer conductors of the cable.
Like $G(x)$, the conductance $G_E(x)$ is essentially a parasitic effect. If the dielectric between the 2 conductors were a perfect insulator, $G_E$ would be zero anywhere in the cable. The imperfections in the dielectric give to $G_E$ finite values randomly distributed along the cable.

The impedance of the entire cable of length $L$ according to this model is given by
\begin{displaymath}
Z_C^{-1}= \int_0^L \left[ G(x) + sC(x) + \frac{s C_E (x)}{1 + s C_E(x) G_E^{-1}(x)} \right] dx=
\end{displaymath}
\begin{equation}
= G_C + s C_C + \int_0^L \frac{s C_E (x)}{1 + s C_E(x) G_E^{-1}(x)} dx.
\label{extimp}
\end{equation}

The result is the same as in the case of the basic model, except for the last term within the integral sign.
This last term is not easily evaluated, unless an assumption about the distribution of $C_E(x)$ and $G_E (x)$ is made.
In \ref{extimp} $C_C$ and $G_C$ would actually be the lower limits for $C_E$ and $G_E$. We prefer to consider them as independent parameters because they are directly measurable.

As in the case of the basic model, the dependence of the impedance values with the position $x$ along the cable accounts for the fact that the cable is not homogenous.
The last term of \ref{extimp} can be simplified with an approximation. Since the capacitance value is mostly due to the geometry of the cable, while the parasitic conductance is due to impurities in the dielectric between the conductors, we expect $C_E(x)$ to be much more homogenous than $G_E(x)$. We can thus drop the dependance of $C_E$ with $x$.
We can also define the time constant $\tau (x)$ = $C_E G_E^{-1}(x)$, and rewrite \ref{extimp} as
\begin{equation}
Z_C^{-1} = G_C + s C_C + s C_E \int_0^L \frac{1}{1 + s \tau(x)} dx.
\label{extimp2}
\end{equation}

We can now make an assumption about the distribution of $\tau (x)$.
In a real situation, $\tau (x)$ would be randomly distributed between two extreme values, defined as $\tau_1$ and $\tau_2$.
To a first order approximation, we can assume it to be uniformely distributed:
\begin{equation}
\tau(x) \simeq \tau_1 + \left(\tau_2-\tau_1\right) \frac{x}{L}.
\label{tauass}
\end{equation}
By writing this, we are also implicitly assuming that the values of $\tau (x)$ are sorted along the cable from $x$ = 0 to $x$ = L. This is not physically reasonable if intended literally. But since the contribution of each element of length $dx$ is independent of the others, we can without loss of generality assume that they could be arranged in an ideal experiment to obtain the case in which the smallest value of $\tau (x)$, which is $\tau_1$, occurs at $x$ = 0, while the largest value, which is $\tau_2$, occurs at $x$ = L.
A similar approach, from which we were inspired, is commonly used to model the statistics of $1/f$ noise in electronic components \cite{vanderziel1, vanderziel2}.

By substituting \ref{tauass} into \ref{extimp2}, we can solve the integral, obtaining
\begin{equation}
Z_C^{-1} = G_C + s C_C + \frac{ C_E }{\tau_2 - \tau_1} \ln \left( \frac{1 + s \tau_2}{1 + s \tau_1} \right).
\label{extimp3}
\end{equation}
The homogeneity condition could be recovered by calculating the limit for $\tau_2 \rightarrow \tau_1$.
But instead, it is more physically reasonable to let $\tau_1 \rightarrow$ 0, meaning that $\tau (x)$ is bound only from above, i.e. there is a minimum value for $G_E (x)$, but not an upper limit (remembering that $C_C$ is the lower limit for the capacitance).
This corresponds to the fact that, for very large values of $G_E$ or above a given frequency, the capacitance $C_C$ dominates the parasitic impedance, and the contribution of $G_E$ and $C_E$ disappears.

\begin{figure}[ht]
\centering
 \includegraphics[scale=0.8]{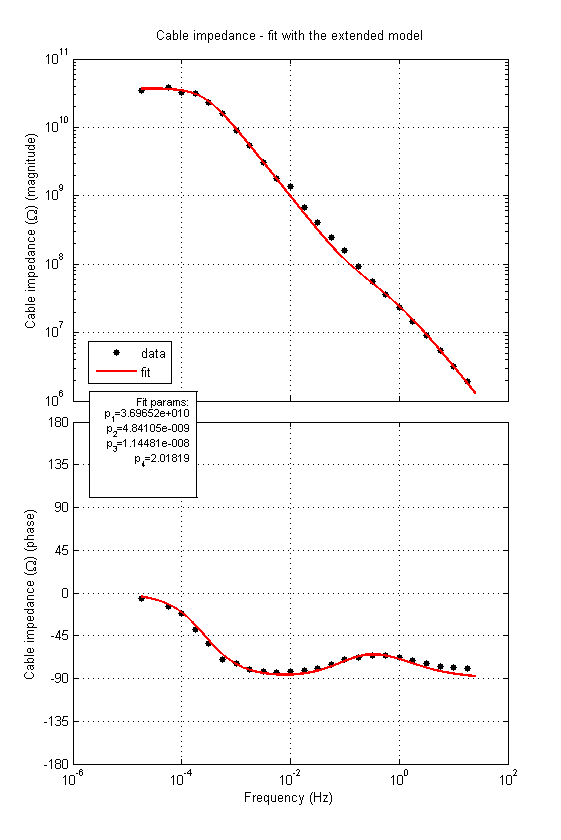}
\caption{The measured cable impedance fitted with the extended model proposed in this paper.}
\label{extendedfit}
\end{figure}

With this last approximation, the inverse of the cable impedance  becomes (also renaming $\tau_2$ as $\tau$):
\begin{equation}
Z_C^{-1} = G_C + s C_C + \frac{ C_E }{\tau} \ln \left( {1 + s \tau} \right).
\label{extimp4}
\end{equation}

The measured cable impedance was then fitted with the following function:
\begin{equation}
f_p = \frac{ p_1 }{1+s p_1 p_2 } || \frac{p_4}{ p_3 \ln \left( 1 + s p_4\right)},
\end{equation} 
where $p_1$ = $G_C^{-1}$ and $p_2$ = $C_C$, as in \ref{fitfunc1}, $p_3$ = $C_E$, $p_4$ = $\tau$, and the symbol $||$ indicates the parallel combination of the two impedances.
The measured data and the best fit with this function are shown in the double plot of figure \ref{extendedfit}. The curve nicely fits the data, both in amplitude and phase. The values obtained from the fit are $G_C^{-1}$=37.0 G$\Omega$, $C_C$ = 4.84 nF, $C_E$ = 11.4 nF and $\tau$ = 2.02 s.
The value obtained from the fit for $C_C$ matches perfectly the value measured at high frequency with a capacitance meter.

This evaluation was carried out also on different cables, finding similar results, and suggesting the general applicability of the extended model.

\afterpage{\clearpage}

\section{Test of the extended model}
\begin{figure}[ht]
\centering
 \includegraphics[scale=0.8]{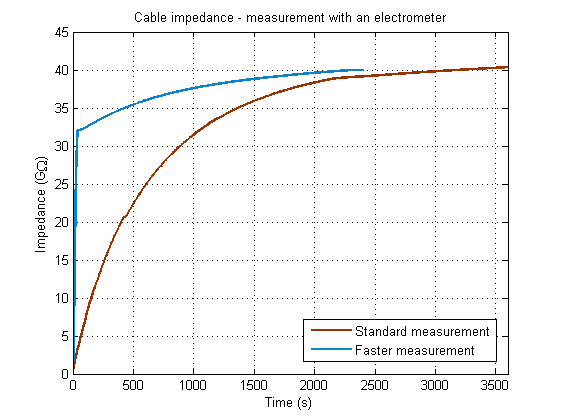}
 \includegraphics[scale=0.8]{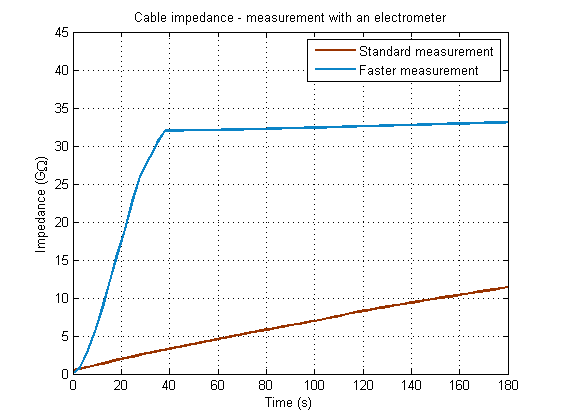}
\caption{The cable impedance versus time, measured with a Keithley 6514 electrometer. The brown curve shows the direct measurement, where the cable impedance was charged with a small constant current for one hour. The blue curve shows the result when a method to speed up the measurement was applied, as described in the text. The lower figure is a detail of the first three minutes of measurement.}
\label{keithley1}
\end{figure}

To check the physical meaningfulness of the model, an independent measurement of $Z_C$ was performed.
A Keithley 6514 electrometer was used for this purpose. The cable was again held at a constant temperature of 30 $^\circ$C in a climatic chamber. When the istrument is measuring the cable parallel resistance $G_C^{-1}$ with the highest sensitivity, up to 200 G$\Omega$, it uses a small current $I$ = 0.9 nA to charge the cable, until the voltage on the cable reaches its final value $V$ = $G_C^{-1}\ I$. The charge is exponential, with a time constant $(C_E + C_C) G_C^{-1}$. The resulting impedance values versus time are shown in the brown curve of figure \ref{keithley1}. By fitting the curve with the exponential function
\begin{equation}
Z_C (t) = \left(1 - e^{-\frac{t}{(C_E + C_C) G_C^{-1}}} \right) G_C^{-1},
\end{equation} 
the values 16.0 nF and 40.1 G$\Omega$ are obtained for $C_E + C_C$ and $G_C^{-1}$ respectively.
The value of $C_E + C_C$ matches the value obtained from the fit of figure \ref{extendedfit}, that is 4.84 + 11.4 = 16.2 nF.
The small discrepancy in the value of $G_C^{-1}$, about 10\%, can be ascribed to the different environmental conditions, most likely a difference in air humidity. This measurement provides thus additional proof of the goodness of the model.

Since all the impedance meaurements presented so far need a large amount of time, a method was also developed to obtain a much quicker estimate of the upper limit for the cable conductance $G_C$.
The digital output of the Keithley 6514 was used to control a relay, which connected a resistance $R_L$ in parallel with the cable.
The value of $R_L$ was chosen as the lower compliance limit for the value of $G_C^{-1}$ divided by 1000, in this case 32 M$\Omega$.
At the beginning of the measurement $R_L$ was put in parallel with the cable. The range of the instrument was set to 200 M$\Omega$, so that the parallel combination of $G_C^{-1}$ and $R_L$ was charged with a constant current of $I$ = 0.9 $\mu$A. The charge time constant is about 1/1000 than in the previous case. The resulting equivalent value for $Z_C$, computed as
\begin{equation}
Z_C = \frac{R_L Z_M}{R_L - Z_M},
\end{equation}
where $Z_M$ is the impedance value given by the Keithley, is shown in the first 40 seconds of the blue curve of figure \ref{keithley1}.
After 40 seconds, when the voltage on the cable was almost settled on the final value, the range of the instrument was changed to 200 G$\Omega$, and the relay was opened, disconnecting $R_L$. From this moment on, the instrument continued to charge the cable with the small current of 0.9 nA, and the charge time constant was the same as for the brown curve.
From the fact that the slope of the measured impedance versus time was still positive after the relay was opened, it is possible to infer that the impedance $G_C^{-1}$ was larger than 32 G$\Omega$.
This faster measurement method allows to give an estimate or an upper limit for $G_C$ in less than three minutes.

\afterpage{\clearpage}

\section{Conclusions}
The inadequacy of the simple basic model, composed of the parallel combination of a resistance and a capacitance to ground, in describing the parasitic impedance of a cable at low frequency was shown. A new model was presented, and its agreement with measurements on samples down to a frequency of 10 $\mu$Hz was proved. While in most practical cases the simple basic model may be enough, the extended model should be adopted for a deeper understanding of the very low frequency behaviour of high impedance connecting links.

\end{document}